\documentstyle[prl,multicol,aps]{revtex}
\begin{document}
\input{epsf.tex}
\epsfverbosetrue

\title{Frequency selection by soliton excitation
in nondegenerate intracavity
downconversion}
\author{Dmitry V. Skryabin$^1$,
Alan R. Champneys$^2$, and William J. Firth$^1$}

\address{$^1$Department of Physics and Applied
Physics, University of
Strathclyde, Glasgow, G4 0NG, Scotland\\
$^2$Department of Engineering Mathematics,
University of Bristol,
Bristol, BS8 1TR, England}

\date{June 14, 1999}

\maketitle

\begin{abstract}
We show that soliton excitation in intracavity
downconversion naturally selects
a strictly  defined frequency difference between
the signal and idler fields.
In particular, this phenomenon implies that if
the signal has smaller losses than the idler
then its frequency is pulled away from the
cavity resonance and the idler frequency
is pulled towards the resonance and {\em vice
versa}.
The frequency selection is shown 
 to be closely linked with the relative
energy balance between the idler and signal
fields.
\end{abstract}

\begin{multicols}{2}
\narrowtext
Exchange of ideas between nonlinear optics and
nonlinear dynamics of spatially distributed
systems has produced a
series of interesting results over the last
decade, and has opened up
one of the most active areas of current
research.
In this particular work we will consider novel
phenomena associated with
soliton excitation in the practically and
fundamentally important area of
 nondegenerate intracavity down-conversion.

The essence of parametric down-conversion is
virtual absorption of a pump
photon at frequency $\omega_p$ with a subsequent
reemission of two photons
with frequencies $\omega_i$ and $\omega_s$,
where indices $i$ and $s$ stand,
respectively, for the idler and signal fields.
Down-conversion can   be realized both in free
propagation and in intracavity schemes.
 The latter takes advantage of positive feedback
provided by the
mirrors and thus transforms the passive free
propagation scheme into an active generator
or optical parametric oscillator (OPO)
\cite{fabre,berger}.

Influence of the transverse degrees of freedom
on the quantum \cite{gl} and classical
\cite{gl,longhipat,stal,degen,domains,longhi,inter,stab,prl}
aspects of the  parametric processes have
recently become a subject of significant
activity. Among the main attraction points  on the
classical side are  localised structures
 \cite{stal,degen,domains,longhi,inter,stab,prl}.
Bright diffractionless localised excitations
inside an optical cavity
supported by different nonlinearities 
({\em cavity solitons}) have  been recently observed
experimentally \cite{weiss} and
suggested for all-optical processing
and storage of information \cite{inter,info},
see also recent reviews \cite{revs}.  
The large quadratic nonlinearities
of artificially phase matched materials
\cite{berger} make parametric
cavity solitons \cite{stal,degen,domains,longhi,inter,stab,prl}
particularly attractive for practical
application, especially where fast material response is an
issue.

Down-conversion processes can  be divided into
degenerate and nondegenerate. In the former case idler and
signal photons are identical while in the latter they differ in
frequencies and/or polarizations.
It has been shown that transverse patterns in
nondegenerate OPO \cite{longhipat,stal}
and soliton dynamics in nondegenerate free
propagation \cite{prl} have qualitative
differences from their degenerate counterparts.
The difference can be formally identified as due to
an additional
symmetry, in the differential phase of signal
and idler fields \cite{prl}.
This symmetry is suppressed in the degenerate
case. As a result the frequency of
the  signal component of any solution, including
solitons, is exactly half the
pump frequency,
$\omega_s\equiv\omega_i=\omega_p/2$.
On the other hand, in the nondegenerate case any
arbitrary
frequency difference $2\Omega$ between the idler
and signal fields still satisfies the condition
$\omega_p=\omega_s+\omega_i$.  This raises the
question of whether there are
any physical constraints on $\Omega$.

While in free propagation the value of $\Omega$
is limited only by
phase matching conditions, this problem becomes
more subtle in the OPO because cavity effects come into
play. A review of early works on this issue,
exploring  approaches based on the
plane wave approximation, can be
found in \cite{fabre}. More recently Longhi \cite{longhipat} has shown
that if diffraction is included
 then $\Omega$ becomes a  function of the
magnitude of the transverse
component  of the signal and idler wave
vectors, $\vec k_{\perp}^i=-\vec k^s_{\perp}$, 
of the exact travelling wave solution which
exists in this system \cite{longhipat}.
For the fixed OPO parameters
$|k_{\perp}^{s,i}|$
can take any values from certain {\em
continuous} bands and hence so can
$\Omega$.

The primary object of this Letter is to
demonstrate that $\Omega$ is
constrained when a cavity soliton is excited in
the nondegenerate OPO.  We
show how this follows from a general relation
between the soliton energies and
the cavity losses.  Existence of this relation,
which has not been previously
identified in this context, seems to be closely
related to survival of {\em differential}
phase symmetry in the presence of cavity
losses.  We use this symmetry to derive
aproximate
formulae for $\Omega$ in certain limits.
Understanding this problem is not only
practically important, but also holds the
key to construction of entire families of cavity
solitons.

Mean-field equations describing interaction of
the signal, idler and pump waves
in OPO \cite{gl,longhipat,longhi} can be
presented in the dimensionless
quasi-Hamiltonian  form
\begin{equation}
(\partial_t+\gamma_m)E_m=i\frac{\delta H}{\delta
E_m^*},~m=s,i,p
\label{eq1}\end{equation}
where $H$ is the following functional:
$H=\int dx[-\alpha_s|\partial_xE_s|^2
-\alpha_i|\partial_xE_i|^2 -
\alpha_p|\partial_xE_p|^2
+\delta_s|E_s|^2+\delta_i|E_i|^2+\delta_p|E_p|^2
+(E_pE_i^*E_s^*+\mu E_s^*E_i^*+c.c.)]$.
$t$ is the  time  measured in the units of
$\tau_cb$, where
$\tau_c$  is the cavity round-trip time and $b$
is an arbitrary scaling
constant. $\gamma_m=T_mb/2$, where $T_m$ are the
effective mirror
transmitivities.
$x=X[2k_s/(bL)]^{1/2}$ and $\alpha_m=k_s/k_m$,
where $X$ is the transverse coordinate in
physical units, $k_m$ are the
longitudinal components  of the wave vectors and
$L$ is a cavity round-trip length.
$\delta_{m}=b\tau_c(\omega_m-\omega_{m}^{cav})$
are the detunings from
the cavity resonances $\omega^{cav}_{m}$.
Physically, validity of these equations requires
small losses and detunings:
$\gamma_m,\delta_m<< b\pi$.
Assuming, for simplicity,  that $\omega_s$ is
close to
$\omega_i$, we can fix
$\alpha_{s,i}=2\alpha_p=1$, but still allow
differences in
$\delta_{s}$, $\delta_{i}$  and $\gamma_{s}$,
$\gamma_{i}$.
Then the phase matching conditions imply
$n(\omega_s)\simeq n(\omega_i)\simeq
n(\omega_p)$, where $n(\omega)$ is the
linear refractive index and the following
scalings  can be derived for the field
variables.
$\mu$ is proportional to the external pump
field  $E_p$:
$E_p=\mu{\sqrt{2(\delta_p^2+\gamma_p^2)/\tilde
T_{\omega_p}}}
/(b\tau_c\chi\omega_s)$,
here  $\chi$ is the effective quadratic
susceptibility.
The physical fields ${\cal E}_m$
are given by
${\cal E}_{s,i}={\sqrt{2}}E_{s,i}e^{i\phi/2}/(b\tau_c\chi\omega_{s})$,
${\cal E}_p=2e^{i\phi}
(E_p+\mu{\sqrt{\delta_p^2+\gamma_p^2}})/(b\tau_c\chi\omega_s)$,
where $\phi=-atan(\gamma_p/\delta_p)$.

We seek localized solutions of Eqs. (1) in the
form
$E_m(x,t)=A_m(x,\Delta)e^{i\Omega_m t}$,
where
$\Omega_{s,i}=\pm\Delta\pm(\delta_s-\delta_i)/2$,
$\Omega_p=0$. Then
$A_m$ obey the set of differential equations
\begin{eqnarray}
\nonumber && -i\gamma_sA_s=(\partial_x^2+\delta-
\Delta)A_s+(A_p+\mu)A_i^*,\\
&& -i\gamma_iA_i=(\partial_x^2+\delta+
\Delta)A_i+(A_p+\mu)A_s^*,\label{eq2}\\
\nonumber &&
-i2\gamma_pA_p=(\partial_x^2+2\delta_p)A_p+2A_sA_i,
\end{eqnarray}
where $\delta=(\delta_s+\delta_i)/2$.
The previously introduced frequency difference
$2\Omega$ between the idler and  signal
fields is linked with the parameter
$\Delta$ by the formula:
$2\Omega=(2\Delta+\delta_s-\delta_i)/(b\tau_c)$
We are interested in bright single-hump
solitons, implying
$A_m(x=+\infty)=0$ and $\partial_x A_m(x=0)=0$.
Existence of such solitons in the parameter
region where the trivial zero solution
bifurcates subcritically can be predicted for $\Delta=0$,
$\gamma_s=\gamma_i$ by analogy with the well studied degenerate
case, where this condition reads
$\delta\delta_p>\gamma_s\gamma_p$
\cite{degen,inter,stab}. In the nondegenerate case solitons have been
found as a result of direct numerical simulations of
Eqs. (\ref{eq1}) starting from either
'random' \cite{stal} or localized \cite{longhi}
initial conditions. Subcritical bifurcation and the related
phenomenon of optical bistability in the nondegenerate OPO
have been demonstrated experimentally in \cite{white96eps}.

For fixed cavity parameters solitons can  exist either for $\Delta$
continuously varying in a certain range or for only a
discrete set of $\Delta$ values. We will show that the latter situation
is realised for  $\gamma_{s,i}\ne 0$.  We thus assert that the cavity
selects the frequency difference between the
signal and idler when a parametric soliton is
excited. Note that the related problem of frequency
selection has been studied in
the general context of the complex
Ginzburg-Landau equation (CGLE) with subcritical
bifurcation \cite{cgle}, which also can be
applied to describe
laser with saturable absorber
\cite{weiss,andrey}. The possibility of the
approximate
reduction of Eqs. (1) to the CGLE  in the
different limits
has been demonstrated in \cite{longhi}.
However,  the problem of the
frequency selection by solitons in OPO  has not
been previously formulated even within the
framework of the CGLE approximation. This fact
has in turn prevented construction of the family
of stationary soliton
solutions of Eqs. (2) and rigorous study of
their stability.

To start our analysis of the frequency
selection, problem, we define the energy
parameters
$Q_m=\int dx |E_m|^2$ and the energy imbalance
$(Q_-=Q_s-Q_i)$. 
Manipulation of Eqs. (1) reveals that the rate
of change of $Q_-$ is given by
\begin{equation}
\partial_tQ_-=-2\gamma_+Q_--2\gamma_-Q_+
=2\gamma_iQ_i-2\gamma_sQ_s,\label{eq3}
\end{equation}
here $\gamma_{\pm}=(\gamma_s\pm\gamma_i)/2$ and
$Q_+=Q_s+Q_i$.
Thus for any steady state solution, such as
soliton solutions of Eqs. (2), the condition
\begin{equation}
\gamma_sQ_s=\gamma_iQ_i,\label{eq4}
\end{equation}
must be satisfied. Eq. (\ref{eq4}) is consistent
with the expectation that
the field with the smaller losses will have the
larger energy.

To further interpret relations (\ref{eq3}), (\ref{eq4})   let us recall that
in free propagation downconversion
is a Hamiltonian process. Then by  Noether's
theorem
every continuous symmetry implies a
corresponding conservation law, see e.g.
\cite{prl}.  Cavity losses  destroy
the Hamiltonian structure  of the problem, see
Eqs. (1), and the input pump breaks
the phase symmetry linked with  conservation of
the total energy $(Q_s+Q_i+2Q_p)$.
The symmetry
$(E_s,E_i)\to(E_se^{i\psi},E_ie^{-i\psi})$ in
the differential phase $\psi$, however,
survives in the cavity configuration, so
how do the losses change the associated law
$\partial_tQ_-=0$?
Self-evidently the relation (\ref{eq3}) becomes
this conservation law in the Hamiltonian limit,
which suggests a more general link between this
energy relation and the symmetry
in the differential phase.

Now using cavity solitons as an example we will
demonstrate that condition
(\ref{eq4}) constrains the frequency
difference of the signal and
idler components of the intracavity field.
We consider signal and idler losses small
compare to the cavity detunings
$\gamma_{s,i}/|\delta|\ll 1$ and set $b=1$.
Conservation of $Q_-$ is restored for
$\gamma_{s,i}=0$ and thus
 Eq. (\ref{eq4}) is  satisfied automatically.
A soliton family then exists for $\delta<0$,
with $\Delta$ {\em continuously}
varying in the interval $(\delta,-\delta)$.
Outside this interval exponential localization
of $A_{s,i}$ is not possible.
Now suppose that for
$\gamma_{s,i}\sim\epsilon\ll 1$  $\Delta$
becomes a slow function of $t$,
i.e. $\partial_t\Delta\sim\epsilon$, then Eq. (3)
immediately gives an equation for the
adiabatic evolution
of $\Delta$:
$\partial_t\Delta\cdot\partial_{\Delta}Q_-
=-\gamma_+Q_-(\Delta)-
\gamma_-Q_+(\Delta)$.
For stationary soliton solutions
$\partial_t\Delta=0$ and
intersections of the curve $Q_s/Q_i$ as a
function of $\Delta$ with the line
$Q_s/Q_i=\gamma_i/\gamma_s$
give selected values of $\Delta$. We plot in
Fig. 1(a) the existence curve
corresponding to the numerically built \cite{numer}
 single-hump
soliton family in the plane $(\mu,\Delta)$  for
$\delta=-1$ and
losses of the signal and idler of order several
percent.
Dots in Fig. 1(a) correspond to values of
$\Delta$ obtained by the
perturbative method.
The agreement is excellent, which also indicates
that the limit of small cavity losses is
non-singular.  The latter is a necessary
condition for a linkage between Eq. (3) and the
differential phase symmetry.
Typical transverse profiles of the soliton
components
are presented in Fig. 1(b).

To study stability of the solitons we seek
solutions of Eqs. (\ref{eq1})
in the form
$(A_m(x)+\epsilon(u_m(x,t)+iw_m(x,t)))e^{i\Omega_mt}$.
After  standard linearization  we derive
$\partial_t\vec\xi=\hat{\cal L}\vec\xi$, where
$\vec\xi=( u_s,u_i,u_p,w_s,w_i,w_p)^T$ and
$\hat{\cal L}$ is the linear non-self-adjoint
differential operator. The discrete spectrum of
$\hat{\cal L}$  has been found numerically
using second-order finite
differences. Any discrete eigenvalue of
$\hat{\cal L}$ with  positive
real part makes the soliton unstable.
The new soliton family turns out to be stable
over the  section $(A,H)$ in Fig. 1(a).
The Hopf instability of the $(H,B)$  branch, and
the instability of the $(O,A)$  branch 
are  similar to the case of the degenerate OPO \cite{stab}.
At the point $B$ the single-hump branch bifurcates back 
in $\mu$, initiating a sequence of multi-hump solitons.

The above perturbation approach to find selected values of
$\Delta$ requires
$\gamma_{s,i}$ small, which is satisfied in most
practical situations, but we also assumed
$\delta\sim O(1)$.   This fails if
$\gamma_{s,i}/|\delta|\sim O(1)$ or $\sim
\epsilon^{-1}$, which physically
means that the cavity becomes tuned close
to resonance with the signal and idler fields.
Then terms proportional to
$\delta$ in Eqs. (2) should also be considered
as perturbations.
In this case Eqs. (2) simply do not have
solitary solutions in zero order  and
therefore $Q_{s,i}$ can't be considered
functions of $\Delta$.
To overcome this difficulty we used another
perturbation approach, also based
on the $\psi$-symmetry.
Note first that the cavity solitons become wider
on approach to resonance  \cite{dijon}, and so
 to avoid large computational windows
it is convenient to take  large $b$, e.g.
$b=2/T_s$.
Then $\gamma_m\sim 1$ physically corresponds to
small losses.
Now, observing that  $A_i=A_s$  is a solution of
Eqs. (2) for $\gamma_s=\gamma_i$ and $\Delta=0$,
we assume $\gamma_-,|\Delta|\sim\epsilon$.
 At  first order in $\epsilon$
\begin{equation}
\epsilon(\hat{\cal
L}-\partial_t)\vec\xi=\Delta\vec\xi_{\psi}
-\gamma_-\vec{\cal P},
\label{eq5}\end{equation} where
$\vec\xi_{\psi}=(-ImA_s,ImA_s,0,ReA_s,-ReA_s,0)^T$
is the neutral mode generated by the
$\psi$-symmetry, i.e. $\hat{\cal L}\vec\xi_{\psi}=0$, and
$\vec{\cal P}=(ReA_s,-ReA_s,0,ImA_s,-ImA_s,0)^T$.
Eq. (5) immediately yields
\begin{equation}
\Delta=\gamma_-\frac{\langle\vec{\cal
P}|\vec\zeta_{\psi} \rangle}
{\langle\vec\xi_{\psi}|\vec\zeta_{\psi}
\rangle},
\label{eq6}\end{equation}
where the new vector $\vec\zeta_{\psi}$ is the
neutral mode of $\hat{\cal L}^{\dagger}$, i.e. 
$\hat{\cal L}^{\dagger}\vec\zeta_{\psi}=0$, 
generated by the
$\psi$-symmetry (which can  be found numerically).
As usual, $\langle..|..\rangle$ defines inner
product in $L_2$.
We  again find  an excellent agreement between
Eq. (\ref{eq6}) and numerical
solutions of Eqs. (2), see Fig. 2(a).
In this case, stability analysis reveals that
the solitons are stable
along the entire segment $(A,B)$, with $B$ again
marking the transition
to multi-hump soliton solutions.
We also  found that $\langle\vec{\cal
P}|\vec\zeta_{\psi} \rangle/
\langle\vec\xi_{\psi}|\vec\zeta_{\psi} \rangle$
is  positive throughout
a wide region of  parameters. This implies that
the sign of $\gamma_-$
determines the sign of $\Delta$, at least for
parameter values where our perturbative approach
is valid. Fig. 1(a) and
selective numerical checks in parameter regions
far outwith the
validity of our perturbative methods support a
conjecture that the relationship
$sgn\gamma_-=sgn\Delta$ has a general character.
Note finally that any difference in  diffraction
constants $\alpha_s$ and $\alpha_i$
will also affect the frequency selection, but
we leave this mechanism for future analysis.

To assess possibilities of experimental
observation of these cavity solitons
let us take as an example a $1cm$ long
monolithic planar waveguide cavity with
$\chi^{(2)}\simeq 20pm/V$,
typical for a noncritically phase matched
potassium niobate crystal.
Suppose the waveguide to be $\sim 1mm$ wide and
$\sim 1\mu m$ thick, excited by an elliptical
pump beam at frequency $\sim 10^{15}Hz$. Other
parameters as for
Fig. 2(a) then imply the following  real world
values: pump power
$\sim \mu^2\times 1W$, selected frequency
difference
$\sim \Omega\times 10^{9}~Hz$ and cavity soliton
width $\sim 30\mu m$. 

To excite cavity solitons one can apply
spatially localised
optical pulses at any of the three frequencies.
Optimisation of the pulse parameters  goes
beyond of our present scope.
However, to demostrate how  the selection of
the frequency difference  happens, we show in
Fig. 3 results
of direct simulation of Eqs. (\ref{eq1}) with,
as initial condition,
a pulse at frequency $\omega_s$ with duration
around  $0.1\tau_c$,  intensity
 several times the peak soliton intensity, and
width of order  the soliton width.
Intensities of all three components of the
excited soliton become constant after a
transient period, see Fig. 3(a),
while the real parts of the signal and idler
fields exhibit undamped oscillations, confirming
selection of $\Delta$
with the predicted value, see Fig. 3(b).

The main physical conclusion which can be drawn
from the above results is that,
if the signal has the smaller losses  its
frequency is pulled
away from the  cavity resonance while the idler
frequency is pulled towards
resonance and {\em vice versa}, see Figs.
1(a),2(a).
If  the  signal and idler losses are equal
then the selected value of $\Delta$ is zero, see
Eq. (\ref{eq6}), which implies that
both the idler and the signal are equally
detuned from cavity resonance.
Thus the cavity structure balances the energies
of the idler and signal
components during soliton excitation, in accord
with Eqs. (\ref{eq3}), (\ref{eq4}).
The understanding of this frequency selection
mechanism has allowed us
to reconstruct an entire  family of single-hump
cavity solitons and to study their stability.
Our results are likely to find applications in
interpretation of
other spatio-temporal phenomena in nondegenerate
OPOs and also to
be relevant in other intracavity parametric
processes with
symmetry in the differential phase, e.g.,
second harmonic generation with competing
parametric process \cite{marte} and
nondegenerate four-wave mixing  \cite{stokes}.

We are  indepted to G.K. Harkness, D. Michaelis
and U. Peschel for  assistance
with numerical problems at the early stage of
the work and to  G.J. de Valc\'arcel for insightful remarks. 
D.V.S. acknowledges financial support from the
Royal Society of Edinburgh and
British Petroleum. The work is partially
supported by
ESPRIT project PIANOS and EPSRC grant GR/M19727.



\begin{figure}
\setlength{\epsfxsize}{9.0cm}
\centerline{\epsfbox{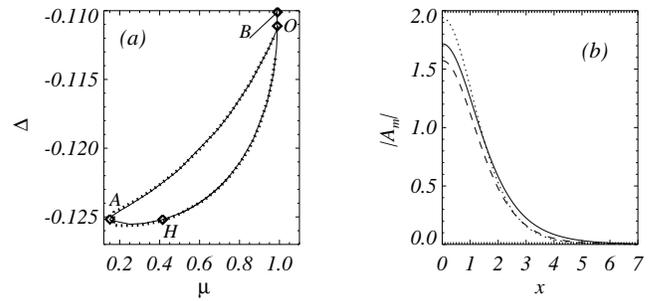}}
\caption{(a) Existence curve in the
$(\mu,\Delta)$-plane for the family of
single-hump cavity solitons, demonstrating
selection of $\Delta$.
Full line corresponds to the results of
numerical continuation, 
dotted line was obtained using perturbative
methods (see text) in the limit
$\gamma_{s,i}/|\delta|\ll 1$:
$\delta=\delta_{p}=-1$, $\gamma_s=0.04$,
$\gamma_i=0.05$, $\gamma_p=0.1$, $b=1$.
Solitons are stable only in the small interval
$(A,H)$, see text for details.
(b) Soliton transverse profile for $\mu=0.3$,
$\Delta=-0.1255007$. Full line:  $|A_s|$,
dashed line: $|A_i|$, dotted line: $|A_p|$.}
\label{fig1}
\end{figure}

\begin{figure}
\setlength{\epsfxsize}{9.0cm}
\centerline{\epsfbox{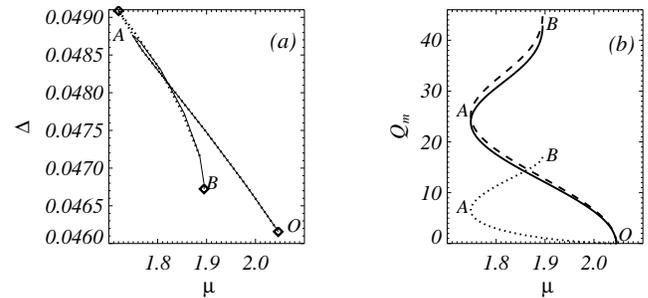}}
\caption{
(a) Existence curve in the $(\mu,\Delta)$-plane
for family of
single-hump cavity solitons, demonstrating
selection of $\Delta$.
Full line corresponds to the results of
numerical continuation, 
dotted line was obtained using Eq. (6):
$\delta=-1.8$, $\delta_{p}=-4$, $\gamma_s=1$,
$\gamma_i=0.95$,
$\gamma_p=2$, $b=2/T_s$. Solitons are stable
only in the interval $(A,B)$.
(b) Plots showing dependence of $Q_m$ vs $\mu$.
Parameters as for (a).
Full line:  $Q_s$, dashed line: $Q_i$, dotted
line: $Q_p$.}
\label{fig2}
\end{figure}

\begin{figure}
\setlength{\epsfxsize}{9.0cm}
\centerline{\epsfbox{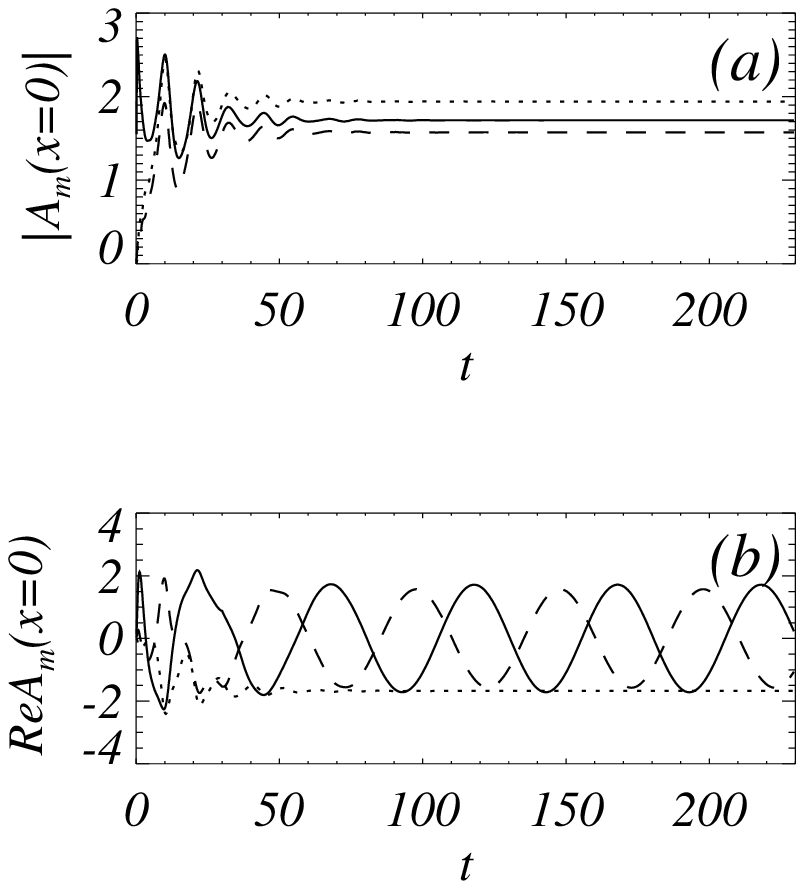}}
\caption{Soliton excitation by a localized
signal-field pulse.
 Temporal evolution of (a) $|E_m|$ and (b)
$ReE_m$ at $x=0$.
Full line: signal, dashed line: idler, dotted
line: pump.
Other parameters as for Fig. 1(b).
Our predicted $\Delta$ corresponds to a period
$\sim 50$ time units.}
\label{fig3}
\end{figure}

\end{multicols}
\end{document}